\documentclass[graybox]{svmult}

\usepackage{mathptmx}       
\usepackage{helvet}         
\usepackage{courier}        
\usepackage{type1cm}        
\usepackage{makeidx}         
\usepackage{graphicx}        
\usepackage{multicol}        
\usepackage[bottom]{footmisc}

\makeindex             

\begin{document}

\title*{Be Stars: Rapidly Rotating Pulsators}
\author{Th.\ Rivinius}
\institute{Th.\ Rivinius \at ESO - European Organisation for Astronomical
    Research in the Southern Hemisphere, Chile \email{triviniu@eso.org}
}
%
%
\maketitle

\abstract*{I will show that Be stars are, without exception, a class of
  rapidly rotating stars, which are in the majority of cases pulsating stars
  as well, while none of them does possess a large scale (i.e.\ with
  significant dipolar contribution) magnetic field.}

\abstract{I will show that Be stars are, without exception, a class of rapidly
  rotating stars, which are in the majority of cases pulsating stars as well,
  while none of them does possess a large scale (i.e.\ with
  significant dipolar contribution) magnetic
  field.}

\section{Introduction}
\label{sec:intro}
A general discussion of classical Be stars and their properties has been
presented by \cite{2003PASP..115.1153P}. In this work, I would like to look
into results obtained since then with respect to the rotational velocity, the
pulsation, and the occurrence of magnetic fields.
 
To define a Be star, a broad working definition is widely employed: ``A
non-supergiant B star whose spectrum has, or had at some time, one or more
Balmer lines in emission'' \cite{1987pbes.coll....3C}. However, it must be
kept in mind that this definition includes a number of object types that have
been awarded their own classes, like interacting binaries of Algol type, stars
in which a magnetosphere clearly dominates the emission and its morphology,
like $\sigma$\,Ori E, and young stars still accreting out of the surrounding
medium or with fossil gaseous disks (for instance 51\,Oph, formerly often
regarded as classical Be star, has been resorted into that group when its
unusual infrared excess became apparent). Therefore, although not useful as a
morphological classification criterion, one should note that apart form
fulfilling the above definition, only such stars are considered classical Be
stars in which this emission arises in a circumstellar disk in Keplerian
rotation (at least its outer parts, see \cite{2011arXiv1111.2487M}), formed by
gas ejected by the star itself.

The question fueling the discussion is the following: What is the mechanism
responsible for ejecting the material from the stellar surface, and adding
sufficient kinetic energy and angular momentum to it to assume Keplerian
rotation, or, as it is usually posed in short: What is the ``Be mechanism''?

There was wide consensus that the rapid rotation of Be stars alone, being at
about 80\% of the critical value ($w=0.8$, defined as
$v_\mathrm{rot}/v_\mathrm{crit}$, with tendency to increase towards later
types), is not sufficient. Therefore, the discussion about 20 years ago
focused on the nature of the short term periodic variations observed both in
spectra and integrated light \cite{1994IAUS..162..311B}, seen in a vast
majority of early Be stars, but (then) rarely or not at all in mid- to
late-type ones. In the next section, I will review that the consensus on
rapid, but not sufficient rotation, has been shaken, and to what extent it
might have been restored, in the one following that there should be little
doubt left that the short term periodic variations are indeed due to
pulsation, and finally that the claim of wide-spread magnetism of Be stars,
contributing to the Be-mechanism, does not hold.

\section{Rotation}
\label{sec:rot}
The consensus of rapid, but not sufficient rotation, increasing towards the
later type Be stars, has been shaken by the interferometric measurement of the
geometrical shape of Achernar \cite{2003A&A...407L..47D}. The discussion
whether or not a residual disk might have influenced the result was settled
with the conclusion that Achernar rotates at $w=0.99$ of its critical value
\cite{2008ApJ...676L..41C}. It has also been suggested that Be star rotational
rates might generally be underestimated, as any value above about $w=0.8$
would not broaden the lines anymore due to gravity darkening, rendering the
equatorial contribution to the line shape insignificant
\cite{2004MNRAS.350..189T}.

Notwithstanding, this question is usually investigated employing statistical
methods on the measured photospheric line widths and shapes, often based on
the compilation by \cite{2001A&A...368..912Y}, like
\cite{2005ApJ...634..585C}, or FASTROT modeling \cite{2005A&A...440..305F},
both including several hundred of stars. However, both approaches come with
problems. One problem is the quality of the input data. For instance, upon
close inspection several Be stars are not single Be stars, but in fact binary
or triple systems where the emission comes from one component, but the
photospheric profiles from another. A famous case was $\beta$\,Cep, for some
time suspected to be an intrinsically slowly rotating (and magnetic) Be star,
unless the binary nature was shown \cite{2006A&A...459L..21S}. However, less
obvious cases exist as well, such as $\nu$\,Gem \cite{2006A&A...459..137R} or
HR\,6819 (Hadrava et al., in prep). Another concern is the influence of the
circumstellar material on the line width and shape. In several cases this has
led to published values of $v\sin i$ differing by hundred and more
\,km\,s$^{-1}$ (like $\zeta$\,Tau, 27\,CMa, $\kappa^1$\,Aps, and
$\mu$\,For). The six examples above were identified in the {\sc Heros}
database of 70 Be stars \cite{2000ASPC..214..356S,2005PAICz..93....1S},
meaning a significant fraction of almost 10\% of a given sample might bias the
result towards a too early-too slow average when relying on traditional $v\sin
i$ statistics.

A more promising approach determines the full set of stellar parameters for a
rapidly rotating star (including inclination and equatorial velocity
separated) not from line width alone, but shape, hoping to eliminate the
gravity darkening problem. They derive $\bar{w}=0.75$
\cite{2005A&A...440..305F}. Apart from that it is not entirely clear how
sensitive the method is to problems like the biases above, a detailed analysis
of the result reveals the certain presence of some own bias: In a histogram of
inclinations, the bin from 80 to 90 degrees is almost empty, while for a
random distribution (which the rest of the histogram follows well) this should
be the most populated one. Also, looking at the inclinations for so-called
shell stars (seen edge on through the disk, i.e.\ the inclination must be
quite equatorial), these are fairly wide distributed between about i=50 and
i=80 degrees, but again depleted for 80 to 90. On the other hand, assuming
$\sin i=1$ for shell stars $\bar{w}\approx0.8$ is obtained for those, but
independent of spectral subtype \cite{2006A&A...459..137R}.

Finally comparing data from the Galaxy, LMC, and SMC a similar value of
$\bar{w}$ as above, for Be star rotation computed back to ZAMS, was found for
the Galaxy, but higher for the LMC and higher still for the SMC, again without
dependency on spectral type \cite{2007A&A...462..683M}.

Putting this all together, there seem to be few things that can be claimed
without debate. One is that there is no single Be star that has been shown to
rotate really slow, i.e.\ below about $w=0.7$ (including uncertainties). As a
population, Be stars seem to have around $\bar{w}\approx 0.8$, but given
potential biases (like above and \cite{2004MNRAS.350..189T}), the example of
Achernar, and the results from SMC vs.\ LMC vs.\ Galaxy, it is more safe to
claim this not to be a mean but rather a lower threshold, that increases as
metalicity decreases. Whether or not a real trend with spectral subtype exists
is unclear.

\section{Pulsation}
\label{sec:puls}
When short term periodic variations were discovered in Be stars, both in
photometry and spectroscopy and with periods typically between 0.5 and 1.5
days, the discussion concentrated on whether these should be understood as
signature of pulsation or some corotating structure
\cite{1994IAUS..162..311B}. With increasing quality of spectroscopic data,
multi-periodicity in distinct mode groups became apparent
\cite{1998A&A...336..177R,2000ASPC..214..232T}, and it was possible to model
the spectroscopic line profile variations as pulsation in great detail
\cite{2003A&A...411..181M,2001A&A...369.1058R} and for a majority of
investigated objects \cite{2003A&A...411..229R}, while the co-rotation
hypothesis, apart from speculation on clouds and differential rotation and a
set of toy models, has not yet produced a quantitative model to reproduce the
multiperiodicity and the line profile variations. With photometric space
missions the multi-periodicity, found in spectroscopic data only with major
efforts, was observed to be the rule in Be stars (e.g.\ Semaan, this volume).

However, this picture is not complete. Again it was first in spectroscopy that
secondary periods were seen, with properties clearly marking them as of
circumstellar origin \cite{1998ASPC..135..348S,2003A&A...411..167S} and
typically about 10\% longer than the photospheric main pulsational periods. In
one intensively studied case, $\mu$\,Cen, these periods were found to last
only for about a dozen of cycles and re-occur with slightly different values
each time \cite{1998A&A...333..125R}. In this case, they were ascribed to
freshly ejected material, undergoing circularization, where the precise
properties (i.e.\ period and phase) depend on the particular properties of the
given ejection. In space photometry of Be stars, often two types of peaks are
observed in the periodogram. One is narrow and easily understood as the
long-term coherent periods corresponding to the stellar pulsation. A second
type is typically much broader than the frequency resolution, and often has
strong, yet variable contribution from harmonics (e.g.\ Semaan, this
volume). The broad nature of these peaks is not consistent with
any corotating feature, but obviously these are the photometric equivalent to
the secondary periods seen in spectroscopy.

\section{Magnetism}
\label{sec:mag}
Claims of large scale magnetic fields in the few ten to few hundred Gauss
regime have been made for a number of Be stars
\cite{2007AN....328.1133H,2009AN....330..708H,2003A&A...409..275N}. Upon close
inspection, with one exception these are in the three to four $\sigma$ regime
of significance, which makes it mandatory to have them confirmed
independently, in particular since most have been claimed with a single
instrument, the reliability of which is debated in the low-significance regime
\cite{2011arXiv1109.5043B,stefano,2011arXiv1109.5436S,2009MNRAS.398.1505S}.
Indeed, so far two of the seven published claims, for $\omega$\,Ori and
$\mu$\,Cen, were firmly rejected by the MiMeS collaboration, with none yet
confirmed \cite{2011arXiv1109.2673G}. A re-reduction of the archival data
comes to an even stronger conclusion: None of these claims hold
\cite{stefano}.

In the MiMeS survey so far 38 Be stars have been probed for the
presence of a magnetic field. Not in a single of them was one found. Comparing
this to the detection statistics of non-emission B stars probed in the same
program (about 8\%, with similar significance limit distribution), the
conclusion from this is that, in any case, Be stars, as a class, are less
magnetic than non-emission B stars.

To this come theoretical problems, as in MHD simulations it was found to be
impossible to form a Keplerian disk through magnetic leverage and release at a
specified distance, for any degree of parameter-tuning, even though the regime
where the angular momentum/kinetic energy balance seems to be correct is well
within reach of MHD modeling (Owocki and ud-Doula, priv. comm.).

\section{Conclusion}
\label{sec:concl}
The problem of the Be mechanism is certainly not solved by the above results,
which are summarized here:

{\bf Rotation:} At this point, the statistical analysis of rotational
velocities of Be stars is not sufficiently robust to base any firm conclusion
on it. A new consensus about Be-star rotation seems to be re-emerging, which
in short might be understood as ``80\% of critical was correct'', but a more
detailed version of it rather reads like ``there is a (metalicity dependent?)
threshold rotation, about 80\% for the MW, of the critical value, below which
no Be stars are formed''.

{\bf Pulsation:} The picture drawn by spectroscopy and space photometry is a
very consistent one: There are long-term coherent photospheric pulsation
periods, in many cases accompanied by circumstellar quasi-periods signaling
the actual mass-transfer from the star to the environment.

{\bf Magnetism:} Not only no magnetic field in any Be star could be confirmed,
but the most thorough search so far, by the MiMeS group, came up with a very
clear and significant null-result, at variance to any other class of
investigated early-type objects.

As a consequence of the above, a Be mechanism relying on (magnetically) forced
corotation on a large scale (i.e.\ other than highly localized) has no
observational support, and as well on theoretical grounds seems to be
excluded.

What is it, then? Rapid rotation must certainly be present. In some cases it
might suffice by itself, yet in many it doesn't. There is no need to require
only one additional mechanism, though. The closer a star rotates to the
critical threshold, the easier it is to launch material into a Keplerian orbit
(tidal forces, weak pulsation, even turbulence). As we retreat from that
limit, less and less mechanisms would get the job done (strong pulsation,
pulsational beating pushing amplitudes well above the sound speed for a
while), until a threshold is reached, below which no mechanism does work any
more. Quite possibly, therefore, the question for the ``Be mechanism'' has to
be recast as the question for the ``Be mechanisms''.

\begin{acknowledgement}
I would particularly like to thank the MiMeS collaboration for permission to
use the results before or in early stages of publication. Discussions with
D.\ Baade, S.\ \v{S}tefl, S.\ Owocki, R.\ Townsend, G.\ Wade, A.\ ud-Doula,
V.\ Petit, M.\ Oksala, C. Neiner, and S.\ Bagnulo have improved this
contribution considerably.
\end{acknowledgement}


\begin{thebibliography}{99.}%
%
%
\bibitem{1994IAUS..162..311B} Baade D., Balona L.~A.:  Periodic Variability 
of Be Stars: Nonradial Pulsation or Rotational Modulation?,  Pulsation; 
Rotation; and Mass Loss in Early-Type Stars,  IAU Symposium 162, 311 (1994)
\bibitem{2011arXiv1109.5043B} Bagnulo S., Landstreet J.~D., Fossati L.,
  Kochukhov O.: Stellar magnetism through the polarized eyes of the FORS1
  instrument, in press in Stellar Polarimetry, ArXiv e-prints, arXiv:1109.5043
  (2011)
\bibitem{stefano} Bagnulo S., Landstreet J.D., Fossati L., Kochukhov O., A\&A,
  in press (2012)
\bibitem{2008ApJ...676L..41C} Carciofi A.~C., Domiciano de Souza A.,
  Magalh{\~a}es A.~M., Bjorkman J.~E., Vakili F.: On the Determination of the
  Rotational Oblateness of Achernar, ApJ \textbf{676}, L41-L44 (2008)
\bibitem{1987pbes.coll....3C} Collins G.~W., II:  The use of terms and 
definitions in the study of Be stars,  IAU Colloq.~92: Physics of Be Stars,  
3-19 (1987) 
\bibitem{2005ApJ...634..585C} Cranmer S.~R.: A Statistical Study of Threshold
  Rotation Rates for the Formation of Disks around Be Stars, ApJ
  \textbf{634}, 585-601 (2005)
\bibitem{2003A&A...407L..47D} Domiciano de Souza A., Kervella P., Jankov S.,
  Abe L., Vakili F., di Folco E., Paresce F.: The spinning-top Be star
  Achernar from VLTI-VINCI, A\&A \textbf{407}, L47-L50 (2003)
\bibitem{2005A&A...440..305F} Fr{\'e}mat Y., Zorec J., Hubert A.-M., Floquet
  M.: Effects of gravitational darkening on the determination of fundamental
  parameters in fast-rotating B-type stars, A\&A
  \textbf{440}, 305-320 (2005)
\bibitem{2011arXiv1109.2673G} Grunhut J.~H., Wade G.~A., the MiMeS
  Collaboration: The incidence of magnetic fields in massive stars: An
  overview of the MiMeS Survey Component, in press in Stellar Polarimetry,
  ArXiv e-prints, arXiv:1109.2673 (2011)
\bibitem{2007AN....328.1133H} Hubrig S., Yudin R.~V., Pogodin M., Sch{\"o}ller
  M., Peters G.~J.: Evidence for weak magnetic fields in early-type emission
  stars, AN \textbf{328}, 1133-(2007)
\bibitem{2009AN....330..708H} Hubrig S., Sch{\"o}ller M., Savanov I., Yudin
  R.~V., Pogodin M.~A., {\v S}tefl S., Rivinius T., Cur{\'e} M.: Magnetic
  survey of emission line B-type stars with FORS 1 at the VLT, AN
  \textbf{330}, 708-(2009)
\bibitem{2003A&A...411..181M} Maintz M., Rivinius T., {\v S}tefl S., Baade D.,
  Wolf B., Townsend R.~H.~D.: Stellar and circumstellar activity of the Be
  star $\omega$\,CMa.  III. Multiline non-radial pulsation modeling, A\&A
  \textbf{411}, 181-191 (2003)
\bibitem{2007A&A...462..683M} Martayan C., Fr{\'e}mat Y., Hubert A.-M.,
  Floquet M., Zorec J., Neiner C.: Effects of metallicity, star-formation
  conditions, and evolution in B and Be stars. II. Small Magellanic Cloud,
  field of NGC\,330, A\&A \textbf{462}, 683-694 (2007)
\bibitem{2011arXiv1111.2487M} Meilland A., Millour F., Kanaan S., Stee P.,
  Petrov R., Hofmann K.-H., Natta A., Perraut K.: First
  spectro-interferometric survey of Be stars I. Observations and constraints
  on the disks geometry and kinematics, in press in A\&A,
  ArXiv e-prints, arXiv:1111.2487 (2011)
\bibitem{2003A&A...409..275N} Neiner C., Hubert A.-M., Fr{\'e}mat Y., Floquet
  M., Jankov S., Preuss O., Henrichs H.~F., Zorec J.: Rotation and magnetic
  field in the Be star $\omega$ Orionis, A\&A \textbf{409}, 275-286 (2003)
\bibitem{2003PASP..115.1153P} Porter, J.~M., Rivinius, T.: Classical Be
  Stars.\ PASP \textbf{115}, 1153-1170 (2003)
\bibitem{2003A&A...411..229R} Rivinius T., Baade D., {\v S}tefl S.:
  Non-radially pulsating Be stars, A\&A \textbf{411}, 229-247
  (2003) \bibitem{1998ASPC..135..348S} {\v S}tefl S., Baade D., Rivinius T.,
  Stahl O., Wolf B., Kaufer A.: Circumstellar Quasi-periods Accompanying
  Stellar Periods of Be Stars, A Half Century of Stellar Pulsation
  Interpretation, 348 (1998)
\bibitem{1998A&A...333..125R} Rivinius T., Baade D., {\v S}tefl S., Stahl O.,
  Wolf B., Kaufer A.: Stellar and circumstellar activity of the Be star
  $\mu$\,Centauri. I. Line emission outbursts, A\&A \textbf{333}, 125-140
  (1998)
\bibitem{1998A&A...336..177R} Rivinius T., Baade D., {\v S}tefl S., Stahl O.,
  Wolf B., Kaufer A.: Stellar and circumstellar activity of the Be star MU
  Centauri. II. Multiperiodic low-order line-profile variability, A\&A
  \textbf{336}, 177-190 (1998)
\bibitem{2001A&A...369.1058R} Rivinius T., Baade D., {\v S}tefl S., Townsend
  R.~H.~D., Stahl O., Wolf B., Kaufer A.: Stellar and circumstellar activity
  of the Be star $\mu$\,Centauri. III. Multiline nonradial pulsation modeling,
  A\&A \textbf{369}, 1058-1077 (2001)
\bibitem{2006A&A...459..137R} Rivinius T., {\v S}tefl S., Baade D.: Bright
  Be-shell stars, A\&A \textbf{459}, 137-145 (2006)
\bibitem{2011arXiv1109.5436S} Shultz M., Wade G., Bagnulo S., Landstreet J.,
  Grunhut J., Hanes D., the MiMeS Collaboration: Magnetic Fields of Slowly
  Pulsating B Stars and $\beta$\,Cep Variables: Comparing Results from FORS1/2
  and ESPaDOnS, , in press in Stellar Polarimetry, ArXiv e-prints,
  arXiv:1109.5436 (2011)
\bibitem{2009MNRAS.398.1505S} Silvester J., Neiner C., Henrichs H.~F., et al.:
  On the incidence of magnetic fields in slowly pulsating B, $\beta$\,Cephei
  and B-type emission-line stars, MNRAS \textbf{398}, 1505-1511 (2009)
\bibitem{2000ASPC..214..356S} {\v S}tefl S., Rivinius T.: HEROS Be Star
  Campaigns, IAU Colloq.~175: The Be Phenomenon in Early-Type Stars, 356
  (2000)
\bibitem{2003A&A...411..167S} {\v S}tefl S., Baade D., Rivinius T., Stahl O.,
  Budovi{\v c}ov{\'a} A., Kaufer A., Maintz M.: Stellar and circumstellar
  activity of the Be star $\omega$\,CMa. II. Periodic line-profile
  variability, A\&A \textbf{411}, 167-180 (2003)
\bibitem{2005PAICz..93....1S} {\v S}tefl S., Rivinius T.: Be star research by
  the HEROS group: results from the past decade, Publications of the
  Astronomical Institute of the Czechoslovak Academy of Sciences \textbf{93},
  1-8 (2005)
\bibitem{2006A&A...459L..21S} Schnerr R.~S., Henrichs H.~F., Oudmaijer R.~D.,
  Telting J.~H.: On the H$\alpha$; emission from the $\beta$ Cephei system,
  A\&A, \textbf{459}, L21-L24 (2006)
\bibitem{2004MNRAS.350..189T} Townsend R.~H.~D., Owocki S.~P., Howarth I.~D.:
  Be-star rotation: how close to critical?, MNRAS \textbf{350}, 189-195
  (2004)
\bibitem{2000ASPC..214..232T} Tubbesing S., Rivinius T., Wolf B., Kaufer 
A.:  Multiperiodic Variability and Outbursts of 28 Cygni,  IAU Colloq.~175: 
The Be Phenomenon in Early-Type Stars, 232 (2000) 
\bibitem{2001A&A...368..912Y} Yudin R.~V.: Statistical analysis of intrinsic
  polarization, IR excess and projected rotational velocity distributions of
  classical Be stars, A\&A \textbf{368}, 912-931 (2001)




\end{thebibliography}
\end{document}